\documentclass[apj]{emulateapj}

\newcommand\bibinc{n}		

\usepackage{subeqnarray}
\usepackage{natbib}
\usepackage{color}
\usepackage[utf8]{inputenc}

\DeclareMathSymbol{\varOmega}{\mathord}{letters}{"0A}
\DeclareMathSymbol{\varSigma}{\mathord}{letters}{"06}
\DeclareMathSymbol{\varPsi}{\mathord}{letters}{"09}

\definecolor{gray}{gray}{0.5}

\begin{document}


\title{The Role of Ice Compositions for Snowlines and the C/N/O Ratios in Active Disks}

\author{Ana-Maria A. Piso\altaffilmark{1,2}, Jamila Pegues\altaffilmark{1,3}, Karin I. \"Oberg\altaffilmark{1}}
\altaffiltext{1}{Harvard-Smithsonian Center for Astrophysics, 60 Garden Street, Cambridge, MA 02138}
\altaffiltext{2}{UCLA, 595 Charles E. Young Drive East, Los Angeles, CA 90095}
\altaffiltext{3}{Department of Astrophysical Sciences, Princeton University, Princeton, NJ 08544}

\begin{abstract}
The elemental compositions of planets define their chemistry, and could potentially be used as beacons for their formation location if the elemental gas and grain ratios of planet birth environments, i.e. protoplanetary disks, are well understood. In disks, the ratios of volatile elements, such as C/O and N/O, are regulated by the abundance of the main C, N, O carriers, their ice binding environment, and the presence of snowlines of major volatiles at different distances from the central star. We explore the effects of disk dynamical processes, molecular compositions and abundances, and ice compositions on the snowline locations of the main C, O and N carriers, and the C/N/O ratios in gas and dust throughout the disk. The gas-phase N/O ratio enhancement in the outer disk (exterior to the H$_2$O snowline) exceeds the C/O ratio enhancement for all reasonable volatile compositions. Ice compositions and disk dynamics individually change the snowline location of N$_2$, the main nitrogen carrier, by a factor of 2-3, and when considered together the range of possible N$_2$ snowline locations is $\sim$11-$\sim$79 AU in a standard disk model. Observations that anchor snowline locations at different stages of planet formation are therefore key to develop C/N/O ratios as a probe of planet formation zones. 


\end{abstract}

\section{Introduction}
\label{sec:intro}

The chemical composition of protoplanetary disks is largely dictated by the freeze-out of volatile species. The snowline locations of volatile molecules are therefore crucial in determining disk chemical abundances in gas and dust, as well as planet compositions.  

Carbon and oxygen bearing molecules, such as H$_2$O, CO$_2$ and CO, as well as the carbon-to-oxygen (C/O) ratio in protoplanetary disks and in giant planet atmospheres have been extensively studied from a theoretical standpoint (\citealt{oberg11}, \citealt{alidib14}, \citealt{madhu14}, \citealt{molliere15}), and snowlines of volatiles such as H$_2$O and CO have been detected (\citealt{zhang13}, \citealt{qi13}). However, disk chemistry involves many other molecular compounds \citep{henning13} including nitrogen bearing species and hydrocarbons (e.g., \citealt{mandell12}), which may affect the compositions of nascent planets.

Both in Solar system comets and in protoplanetary disks, volatile carbon and oxygen are primarily contained in H$_2$O, CO$_2$ and CO (e.g., \citealt{lodders03}, \citealt{mumma11}, \citealt{oberg11}, \citealt{boogert15}). However, some fraction of carbon may also be carried by CH$_4$ (e.g., \citealt{oberg08}), which may change the C/O ratio in gas and in dust at some disk locations. In the case of nitrogen, chemical models of the protostellar nebula (e.g., \citealt{owen01}) and of protoplanetary disks (e.g., \citealt{rodgers02}) suggest that N$_2$ was the dominant form of nitrogen, and that giant planets have accreted their nitrogen content primarily as N$_2$ \citep{mousis14}. 
Due to the high volatility of N$_2$, the gas phase nitrogen-to-oxygen (N/O) ratio in the outer disk is expected to be high, perhaps more enhanced than the C/O ratio compared to the Solar value. 
The C/O ratio in gas cannot exceed unity (i.e., a factor of $\sim$2 enhancement compared to the Solar value) since the major volatile carbon carrier is CO. In contrast, the N/O ratio mainly depends on the relative depletion of N$_2$ and oxygen carriers, and it will increase as each of the oxygen carrier snowlines (H$_2$O, CO$_2$, CO) is crossed. Beyond the CO snowline, there is no strict upper limit to the N/O ratio. The spatial extent of this latter region depends on the relative bond strengths of CO and N$_2$ to ice, but may be quite large (see Section \ref{sec:results}). Giant planets that form at wide separations should thus have an excess of nitrogen in their atmospheres, which could be used to trace their formation origin. In addition to N$_2$, a fraction of the nitrogen abundance may also be carried by less volatile species such as NH$_3$ (\citealt{bottinelli10}, \citealt{mumma11}). The present day N$_2$ in Titan's atmosphere, for example, is thought to originate from accretion of primordial NH$_3$ (\citealt{atreya78}, \citealt{mandt14}).

The snowline locations of the main carbon, oxygen and nitrogen carriers strongly depend on the ice grain composition. Very volatile species, such as CO and N$_2$, present binding energies, and therefore snowline locations, that are sensitive to the details of the composition of the icy grain mantles. 
Spectroscopic observations suggest that up to 90\% of the CO ice is frozen in a layer that is thick enough and separated from the H$_2$O ice layer underneath so that it can be considered pure ice (e.g., \citealt{pontoppidan03}). However, if the CO (or N$_2$) ice layer is thin enough ($\sim$monolayer coverage), then it will interact with the H$_2$O ice substrate (e.g., \citealt{collings03}). The ice binding energy is significantly larger in this water dominated environment than in the pure ice case, as shown by laboratory experiments (\citealt{collings03}, \citealt{oberg05}, \citealt{bisschop06}, \citealt{fayolle16}). 
This implies that ices in different environments will sublimate at different radii, which will substantially change the disk regions where these volatiles are present in gaseous or solid form (see Section \ref{sec:static}).  In protostellar cores, H$_2$O ice is primarily amorphous (e.g., \citealt{williams02}, \citealt{vandishoeck14}).  When temperatures exceed $\sim$80-90 K, H$_2$O ice acquires a crystalline structure \citep{schegerer10}. The CO binding energy is larger in an amorphous porous H$_2$O ice environment than in the amorphous compact or crystalline cases (\citealt{noble12}, \citealt{fayolle16}). The N$_2$ binding energy is also larger for a porous versus compact H$_2$O ice substrate. No equivalent studies have yet been performed for the deposition of N$_2$ on crystalline H$_2$O ice, but we expect the N$_2$ binding energy in this environment to follow the same trend as the CO binding energy, as CO and N$_2$ display a similar desorption behavior. To explore the range of distances at which CO and N$_2$ in different environments desorb, we consider the limiting cases: pure ices (lowest binding energy) and ices residing on an amorphous porous H$_2$O ice substrate (highest binding energy). We refer to the latter simply as water dominated ices unless noted otherwise.

 


In this work, we expand the coupled drift-desorption model developed in (\citealt{piso15b}; hereafter Paper I) by considering additional volatile molecules and abundances, ice compositions, as well as nitrogen-to-oxygen (N/O) ratios. This paper is organized as follows. In Section \ref{sec:review}, we review the drift-desorption model developed in Paper I. We discuss the effect of different abundances of the main carbon, oxygen and nitrogen carriers, grain compositions and disk dynamics on snowline locations and the C/N/O ratios in Section \ref{sec:results}. We address the implications of our results in Section \ref{sec:discussion} and summarize our findings in Section \ref{sec:summary}.  



\section{Coupled Drift-Desorption Model}
\label{sec:review}

We begin with a brief review of Paper I's model for the effect of radial drift and viscous gas accretion on volatile snowline locations. We review our disk model in Section \ref{sec:disktime}, and summarize our numerical method and results in Section \ref{sec:driftdes}.

\subsection{Disk Model}
\label{sec:disktime}

In this work we consider both a static and a viscous disk. The static disk is irradiated by the central star and does not experience redistribution of solids or radial movement of the nebular gas. To quantify the effects of radial drift and gas accretion, we use a viscous disk with a spatially and temporally constant mass flux, $\dot{M}$. The viscous disk takes into account radial drift, gas accretion onto the central star, as well as accretion heating. We focus on this disk model which includes all the dynamical and thermal processes we are interested in for the scope of this paper, and do not further consider the other disk models presented in Paper I.  

Following \citet{chiang10}, the temperature profile for a static disk is
\begin{equation}
\label{eq:diskT}
T_{\rm irr} = 120\, (r/\text{AU})^{-3/7} \,\,\text{K},
\end{equation}
where $r$ is the semimajor axis. We use the \citet{shakura73}  steady-state disk solution to model the viscous disk. From Paper I, the viscous disk temperature profile is computed as 
\begin{equation}
\label{eq:activeT}
T_{\rm visc}^4 =\Big[\frac{1}{4 r} \Big(\frac{3 G \kappa_0\dot{M}^2 M_* \mu m_{\rm p} \Omega_{\rm k}}{\pi^2 \alpha k_{\rm B} \sigma}\Big)^{1/3}\Big]^4 + T_{\rm irr}^4.
\end{equation}
Here $G$ is the gravitational constant, $\kappa_0=2 \times 10^{-6}$ is a dimensionless opacity coefficient, $M_*=M_{\odot}$ is the mass of the central star, $\mu=2.35$ is the mean molecular weight of the nebular gas, $m_{\rm p}$ is the proton mass, $\Omega_{\rm k}=\sqrt{G M_{\odot}/r^3}$ is the Keplerian angular velocity, $\alpha=0.01$ is a dimensionless coefficient (see below for details), $k_{\rm B}$ is the Boltzmann constant, and $\sigma$ is the Stefan-Boltzmann constant. 


The steady-state disk has an $\alpha$-viscosity prescription, where the kinematic viscosity is $\nu=\alpha c H$. Here $c \equiv \sqrt{k_{\rm B} T_{\rm visc} /(\mu m_{\rm p})}$ is the isothermal sound speed and $H \equiv c/\Omega_{\rm k}$ is the disk scale height. We can then determine the gas surface density for a viscous disk as (\citealt{shakura73}; see also Paper I for a more detailed explanation of these calculations):
\begin{equation}
\label{eq:Sigmaact}
\Sigma=\frac{\dot{M}}{3 \pi \nu}.
\end{equation}
We choose $\dot{M}=10^{-8} M_{\odot}$ yr$^{-1}$, consistent with mass flux observations in disks (e.g., \citealt{andrews10}). As described in Paper I, the mass flux rate $\dot{M}$ and stellar luminosity $L_*$ will vary throughout the disk lifetime (\citealt{kennedy06}, \citealt{chambers09}), in contrast with our simplified model which assumes that both quantities are constant. This effect will be most pronounced in the inner disk ($\lesssim$ few AU), where accretion heating dominates. We thus acknowledge that the location of the H$_2$O snowline may be determined by the decline in $\dot{M}$ or $L_*$ with time, rather than radial drift (see Paper I, Section 2.1 for a more detailed explanation). 

%

\subsection{Desorption-Drift Equations and Results}
\label{sec:driftdes}

\begin{figure*}[t!]
\centering
\includegraphics[width=0.95\textwidth]{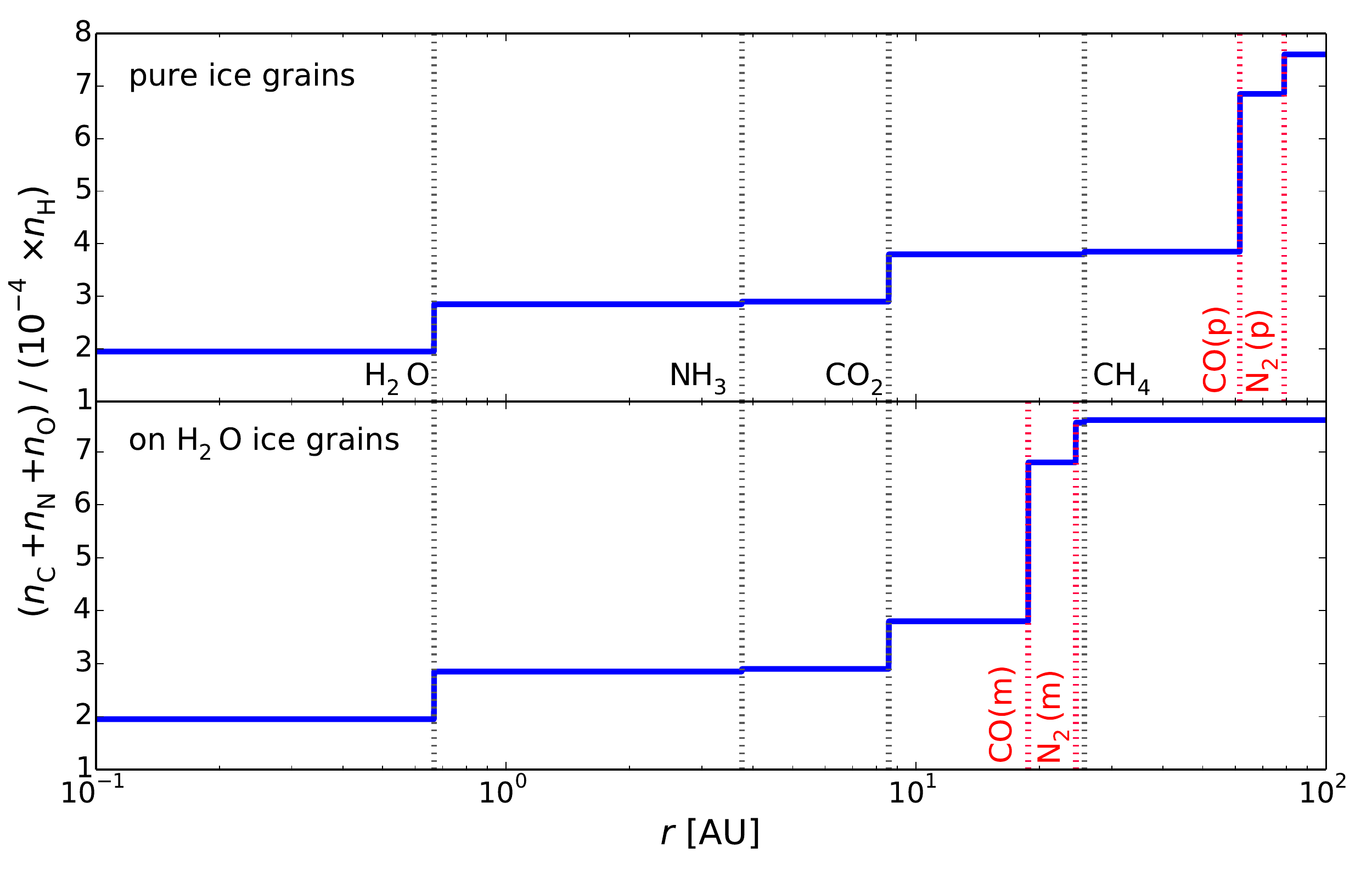}
\caption{The total carbon, nitrogen and oxygen abundance in solids as a function of semimajor axis in a static disk, for CO and N$_2$ as pure ices (top panel) and water dominated ices (bottom panel). Relevant volatile snowlines are marked by the vertical dashed lines. The grain abundances are calculated as a function of the observed median CH$_4$ and NH$_3$ abundances in protostellar cores. The total grain abundance increases with semimajor axis as more and more species freeze out.} 
\label{fig:CNOstatic}
\end{figure*}

The model is described in full in Paper I, here we review and summarize key concepts and results. For a range of initial icy grain sizes composed of a single volatile, we showed in Paper I that the timescale on which these particles desorb is comparable to their radial drift time, as well as to the accretion timescale of the nebular gas onto the central star. We thus have to take into account both drift and gas accretion when we calculate the disk location at which a particle desorbs, since that location may be different from the snowline position in a static disk for a given volatile (see Figure \ref{fig:CNOstatic} and \citealt{oberg11}). We determine a particle's final location in the disk by solving the following coupled differential equations:
\begin{subeqnarray}
\label{eq:ddt}
\frac{ds}{dt} &= & - \frac{3 \mu_x m_{\rm p}}{\rho_{\rm s}} N_x R_{\rm des, x}  \slabel{eq:dsdt} \\
\frac{dr}{dt} &=& \dot{r} \slabel{eq:drdt},
\end{subeqnarray}
where $s$ is the particle size, $t$ is time, $\mu_x$ is the mean molecular weight of volatile $x$, $\rho_{\rm s}=2$ g cm $^{-3}$ is the density of an icy particle, $N_x \approx 10^{15}$ sites cm $^{-2}$ is the number of adsorption sites of molecule $x$ per cm$^{-2}$, $R_{\rm des, x}$ is the desorption rate of species $x$, and $\dot{r}$ is the particle's radial drift velocity. We calculate $R_{\rm, des}$ and $\dot{r}$ as follows.

The desorption rate $R_{\rm des, x}$ (per molecule) is \citep{hollenbach09}
\begin{equation}
\label{eq:Rdes}
R_{\rm{des}, x} = \nu_x \exp{(-E_x/T_{\rm grain})},
\end{equation}
where $E_x$ is the adsorption binding energy in units of Kelvin, $T_{\rm grain}$ is the grain temperature (assumed to be the same as the disk temperature, see Paper I), and $\nu_x=1.6 \times 10^{11} \sqrt{(E_x/\mu_x)}$ s$^{-1}$ is the molecule's vibrational frequency in the surface potential well. We discuss our choices for $E_x$ for the different volatile species in Section \ref{sec:snowlines}.

Following \citet{chiang10} and \citet{birnstiel12}, a particle's radial drift velocity can be approximated as 
\begin{equation}
\label{eq:rdotact}
\dot{r} \approx -2 \eta \Omega_{\rm k} r \Big(\frac{\tau_{\rm s}}{1+\tau_{\rm s}^2}\Big) + \frac{\dot{r}_{\rm gas}}{1+\tau_{\rm s}^2},
\end{equation}
where the first term is the drift velocity in a non-accreting disk and the second term accounts for the radial movement of the gas. Here $\eta \approx c^2/(2 v_{\rm k}^2)$, where $v_{\rm k}$ is the Keplerian velocity, and $\tau_{\rm s} \equiv \Omega_{\rm k} t_{\rm s}$ is the dimensionless stopping time:
\begin{equation}
\label{eq:ts}
t_{\rm s}= \left\{
\begin{array}{l l}
\rho_{\rm s} s / (\rho c), & \quad s < 9 \lambda/4 \,\,\,\ \text{Epstein drag} \\
4 \rho_{\rm s} s^2 / (9 \rho c \lambda), & \quad s < 9 \lambda/4, \,\text{Re} \lesssim 1 \,\,\,\ \text{Stokes drag,}
\end{array} 
\right.
\end{equation}
where $\rho$ is the disk mid-plane density, $\lambda$ is the mean free path and Re is the Reynolds number.  The gas accretion velocity $\dot{r}_{\rm gas}$ is determined from $\dot{M}=-2 \pi r \dot{r}_{\rm gas} \Sigma$, for a fixed $\dot{M}$ and with $\Sigma$ given by Equation (\ref{eq:Sigmaact}). 

For a particle of initial size $s_0$, we solve the Equation set (\ref{eq:ddt}) with the initial conditions $s(t_0)=s_0$ and $r(t_0)=r_0$, where $t_0$ is the time at which we start the integration and $r_0$ is the particle's initial location. We stop our simulation after $t_{\rm d}=3$ Myr, the disk lifetime, since this is roughly the timescale on which planets form, and determine the desorption timescale $t_{\rm des}$ from $s(t_{\rm des})=0$, and thus a particle's desorption distance $r_{\rm des}=r(t_{\rm des})$. Our results are insensitive to our choice of $t_0$ as long as $t_0 \ll t_{\rm d}$. We note that a particle's size is initially fixed and only changes due to desorption. We thus do not take into account processes such as grain coagulation or fragmentation, which nonetheless occur in disks (e.g., \citealt{birnstiel12}, \citealt{perez12}). We discuss the effect of these processes on snowline locations in Paper I.

As we show in Paper I, a particle of initial size $s_0$ can experience three outcomes after $t_{\rm d}=3$ Myr: (1) it can remain at its initial location, (2) it can drift towards the host star, then stop without evaporating significantly, and (3) it can completely desorb on a timescale shorter than 3 Myr. Particles in scenarios (1) and (2) are thus not affected by radial drift or gas accretion, and the snowline locations are those for a static disk. In contrast, the grains in case (3) desorb practically \textit{instantaneously} and \textit{at a fixed particle-size dependent location} in the disk, regardless of their initial position. The snowline locations for these particles will thus be fixed for a given initial particle size and disk model. We have found that grains with sizes $\sim0.001$ cm $\lesssim s \lesssim$ 7 m satisfy this condition for our fiducial disk. 


\section{Results}
\label{sec:results}

\subsection{Snowlines in a Static Disk: The Importance of Ice Compositions}
\label{sec:snowlines}

As we note in Section \ref{sec:intro}, the disk volatile composition and the ice composition determine the location of important snowlines. In this work we focus on the primary carbon, oxygen and nitrogen carriers, i.e. H$_2$O, CO$_2$, CO, N$_2$, and to a lesser extent, CH$_4$ and NH$_3$. 
Our standard model is based on the median ice abundances observed toward Solar-type protostars \citep{oberg11a}, which are $n_{\rm CO_2}=0.29 \times n_{\rm H_2O}$, $n_{\rm CO}=0.38 \times n_{\rm H_2O}$, $n_{\rm CH_4}=0.0555 \times n_{\rm H_2O}$ (hereafter CH$_4$-mid) and $n_{\rm NH_3}=0.055 \times n_{\rm H_2O}$ (hereafter NH$_3$-mid). Here $n_{\rm H_2O} \approx 10^{-4} \times n_{\rm H}$ is the total water abundance \citep{vandishoeck06}, with $n_{\rm H}$ the hydrogen abundance in the disk midplane. For CO, we also take into account that the observed CO ice only traces some of the CO reservoir due to its high volatility, and similarly to \citet{oberg11} and Paper I we set the total CO abundance to $0.9 \times 10^{-4} n_{\rm H}$. Finally, we assume that all nitrogen not found in NH$_3$ is in N$_2$ and assume a Solar nitrogen abundance, $n_{\rm N}=8 \times 10^{-5} n_{\rm H}$ \citep{lodders03}. In effect, this model assumes no chemical evolution between the protostellar and disk midplane stages. This is reasonable for material that accretes onto the disk at large radii \citep{visser09}, but may overestimate the contribution of the original volatiles to the total volatile budget in the innermost disk. 

We determine the location of the H$_2$O, CO$_2$, CO, CH$_4$, N$_2$ and NH$_3$ snowlines in our static disk by balancing desorption with readsorption, following \citet{hollenbach09}. The binding energies of H$_2$O, CO$_2$, CO, CH$_4$, N$_2$ and NH$_3$ as pure ices are 5800 K, 2000 K, 834 K, 1300 K, 767 K and 2965 K, respectively (\citealt{fraser01}, \citealt{collings04}, \citealt{fayolle16}, \citealt{garrod06}, \citealt{martin14}). For our fiducial disk model, these energies correspond to disk temperatures of 143 K, 48 K, 21 K, 30 K, 18 K and 68 K, respectively. For CO and N$_2$ as water dominated ices, the binding energies are 1388 K and 1266 K, respectively \citep{fayolle16}, corresponding to disk temperatures of 34 K and 31 K, respectively. Figure \ref{fig:CNOstatic} shows the resulting snowline locations, assuming CO and N$_2$ pure ices (top panel), and CO and N$_2$ in water dominated ices (bottom panel). The ordinate displays the total carbon, oxygen and nitrogen abundance in solids as a function of the hydrogen total abundance.  As expected, the total grain abundance increases with semimajor axis, as more and more species freeze out. Freeze-out at the CO$_2$ and CO snowlines pulls more heavy elements into the grains than in the case of the H$_2$O snowline.  

Figure \ref{fig:CNOstatic}, bottom panel, displays the snowline locations when CO and N$_2$ are in an amorphous porous water environment (see Section \ref{sec:intro}).  The snowlines move outward by a factor of $\sim$3.3 if the ices are pure, and by up to a factor of $\sim$2 for an amorphous compact water substrate (not shown; our estimates are based on the results of \citealt{fayolle16}). The CO snowline moves outward by a factor of $\sim$1.5 if CO is in a crystalline rather than an amorphous porous water environment (not shown; our estimates are based on the results of \citealt{noble12}), and a similar trend is expected for the N$_2$ snowline (see Section \ref{sec:intro}). 
The results of Figure \ref{fig:CNOstatic} thus represent the limiting cases for the positions of the CO and N$_2$ snowlines, for different compositions of the icy grains and morphology of the water ice substrate. This variation in snowline locations changes the chemical abundances both in gas and dust throughout the disk, directly affecting the compositions of nascent giant planets forming in situ. 
In our simple model, we ignore the effects of CO and N$_2$ entrapment in water ice through clathrate formation or other processes. Theoretical models aimed at explaining the composition of comet 67P/Churyumov-Gerasimenko suggest that a small fraction of the total CO and N$_2$ reservoir may be trapped in clathrates, and only released upon water sublimation (\citealt{lectez15}, \citealt{mousis16}). In this case, the CO and N$_2$ snowlines would be closer to the star than in the pure ice case.


\subsection{C/N/O Ratios in Static Disks}
\label{sec:static}

In this section we determine the C/O and N/O ratios in gas and dust throughout our static disk, and to what extent they are affected by the presence of CH$_4$ and NH$_3$ over the full range of observed CH$_4$ and NH$_3$ abundances toward low-mass protostars. In this section we only consider pure ices, i.e. ices that are layered on a silicate mantle. In reality, the icy grain will have a layered structure with volatiles residing on top of a H$_2$O ice substrate. However, if the volatile ice layer is thick enough and separated from the H$_2$O ice layer, the interaction between H$_2$O and the other volatile species will be minimal and thus the ices can be considered pure rather than water dominated (see also Section \ref{sec:intro}).


We explore the parameter space of possible CH$_4$ abundances by assuming three different scenarios: (1) no CH$_4$, (2) CH$_4$-mid, and (3) the maximum CH$_4$ observed abundance (hereafter CH$_4$-max), $n_{\rm CH_4-max}=0.13 \times n_{\rm H_2O}$ \citep{oberg08}. 
Since the abundance of carbon grains is uncertain, we assume that all the carbon that is not in the form of CH$_4$, CO and CO$_2$ is in carbon grains, so that we reproduce the Solar C/O ratio (gas+dust) of 0.54.  

\begin{figure}[h!]
\centering
\includegraphics[width=0.5\textwidth]{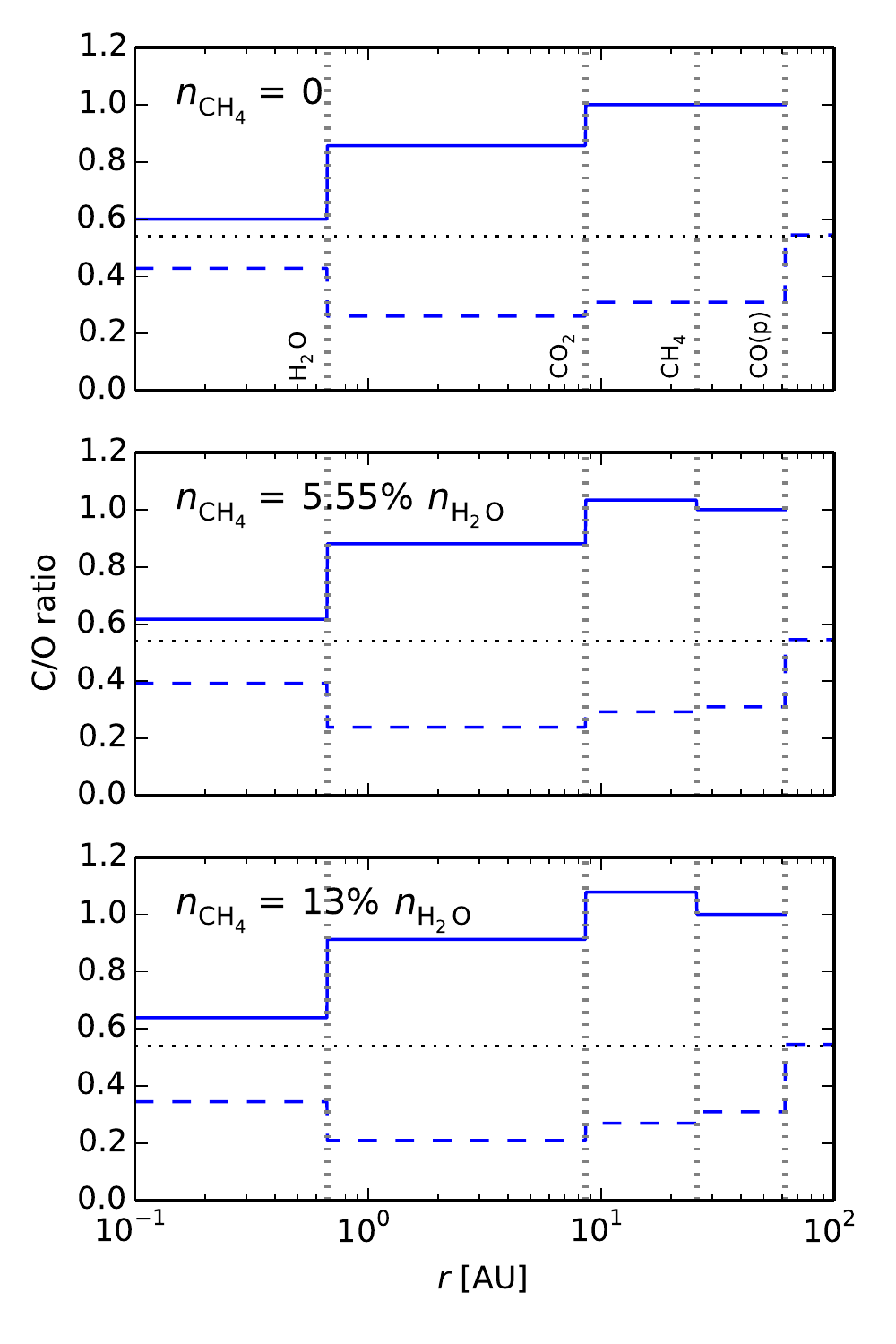}
\caption{The C/O ratio in gas (solid lines) and dust (dashed lines) as a function of semimajor axis in a static disk, assuming no carbon is present in the form of CH$_4$ (top panel), the median observed CH$_4$ abundance is assumed (middle panel), and the maximum observed CH$_4$ abundance is assumed (bottom panel). The C/O estimates are performed assuming that the CO ices are in pure form. The vertical dotted lines mark the snowline locations of the main C and O carriers. The horizontal dotted lines represent the stellar C/O value. The presence of methane only modestly increases the C/O ratio in gas between the CO$_2$ and CH$_4$ snowlines.} 
\label{fig:COstatic}
\end{figure}

Figure \ref{fig:COstatic} shows the C/O ratio in gas and dust as a function of semimajor axis in a static disk: no CH$_4$ (top panel), CH$_4$-mid (middle panel) and CH$_4$-max (bottom panel). As in \citet{oberg11} and Paper I, a gaseous C/O ratio of unity can be achieved between the CO$_2$ and CO snowlines, where oxygen gas is significantly depleted. The gas-phase C/O ratio may be further enhanced between the CO$_2$ and CH$_4$ snowlines due to the presence of additional carbon gas from CH$_4$. In this region, the C/O ratio increases by 3\% for CH$_4$-mid and by 8\% for CH$_4$-max, as displayed in the middle and bottom panels of Figure \ref{fig:COstatic}. Based on the range of observed CH$_4$ protostellar abundances, its presence in the disk only modestly affects the C/O ratio. 

We assume that the main nitrogen-bearing species are N$_2$ and NH$_3$, since other volatiles that contain nitrogen have significantly lower abundances in comparison (e.g., \citealt{mumma11}). Similarly to the case of CH$_4$, we explore the parameter space of possible NH$_3$ abundances using observations toward low-mass protostars, as follows: (1) no NH$_3$, (2) NH$_3$-mid, and (3) the maximum observed NH$_3$ abundance $n_{\rm NH_3-max}=0.15 \times n_{\rm H_2O}$ \citep{bottinelli10}. In each case, the N$_2$ abundance then simply follows as $n_{\rm N_2}=(n_{\rm N}-n_{\rm NH_3})/2$.

\begin{figure}[h!]
\centering
\includegraphics[width=0.5\textwidth]{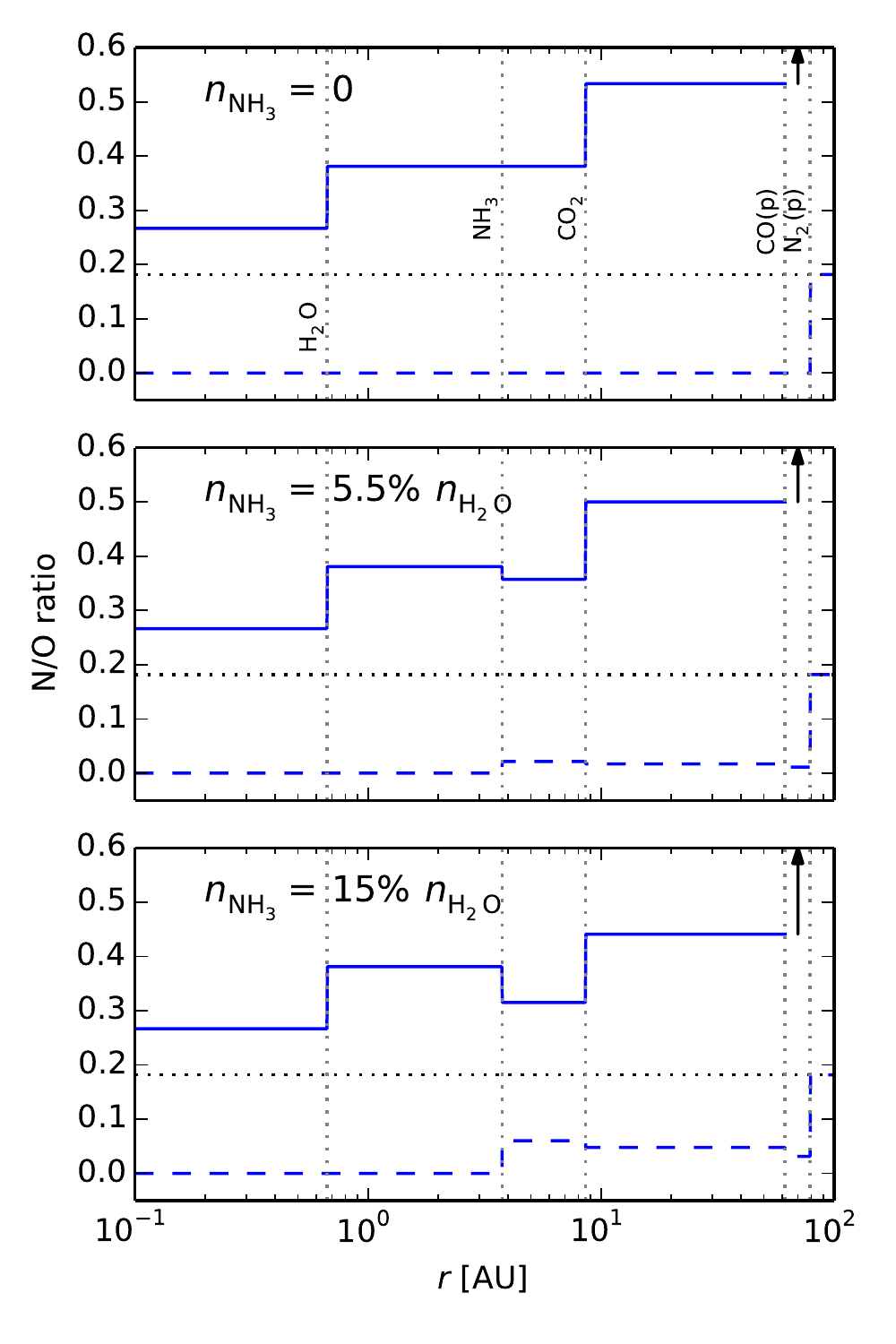}
\caption{The N/O ratio in gas (solid lines) and dust (dashed lines) as a function of semimajor axis in a static disk, assuming no nitrogen is present in the form of NH$_3$ (top panel), the median observed NH$_3$ abundance is assumed (middle panel), and the maximum observed NH$_3$ abundance is assumed (bottom panel). The N/O estimates are performed assuming that the CO and N$_2$ ices are in pure form. The vertical dotted lines mark the snowline locations of the main C,O and N carriers. The horizontal dotted lines represent the average N/O value in the disk. The gas-phase N/O ratio is enhanced by a factor of two between the H$_2$O and CO$_2$ snowlines compared to its average value, and by a factor of three between the CO$_2$ and CO snowlines. The arrows mark a highly elevated N/O ratio in gas between the CO and N$_2$ snowlines due to the depletion of oxygen gas in this region. The presence of NH$_3$ moderately decreases the N/O ratio in gas between the NH$_3$ and CO$_2$ snowlines.} 
\label{fig:Nstatic}
\end{figure}

Figure \ref{fig:Nstatic} shows the snowline locations of the main oxygen and nitrogen carriers and the N/O ratio in gas and dust as a function of semimajor axis in a static disk, for our three choices of the NH$_3$ abundance: no NH$_3$ (top panel), NH$_3$-mid (middle panel) and NH$_3$-max (bottom panel). For comparison, the horizontal dotted lines show the average N/O ratio in the disk. As expected, the gaseous N/O ratio generally exhibits an increasing trend towards the outer disk as more oxygen gas is depleted, with small decreases between the NH$_3$ and CO$_2$ snowlines (by 6\% for NH$_3$-mid and by 18\% for NH$_3$-max, respectively) due to NH$_3$ freeze-out. While the presence of NH$_3$ only moderately affects our results for the N/O ratio, NH$_3$ is important since otherwise the nitrogen content in solid bodies would be more depleted than is observed for comets and asteroids (\citealt{wyckoff91}, \citealt{mumma11}, \citealt{bergin15}).

The gas-phase N/O ratio is enhanced by a factor of two outside the H$_2$O snowline compared to its average value, by more than a factor of three between the CO$_2$ and CO snowlines, and by orders of magnitude between the CO and N$_2$ snowlines. This latter region can span tens of AU depending on disk parameters and the relative CO and N$_2$ ice binding environment. This N/O  enhancement is more pronounced than the C/O gas phase enhancement of a factor of two in the outer disk (see Figure \ref{fig:COstatic}). 

\subsection{C/N/O Ratios in Dynamic Disks}
\label{sec:dynamic}

\begin{figure}[h!]
\centering
\includegraphics[width=0.5\textwidth]{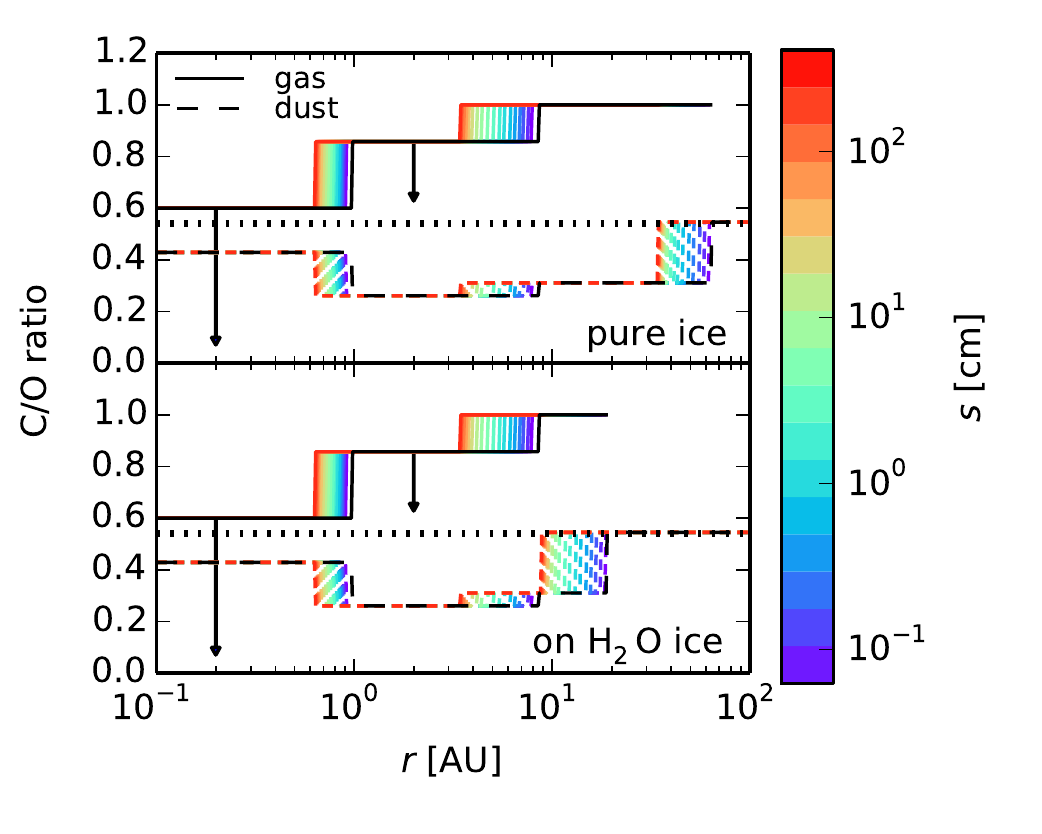}
\caption{C/O ratio estimates in gas (solid lines) and dust (dashed lines) as function of semimajor axis in a viscous disk, for CO as pure ice (top panel) or as water dominated ices (bottom panel). The H$_2$O, CO$_2$ and CO snowlines are shown for particles with initial sizes $\sim0.05$ cm $\lesssim s \lesssim$ 7 m as indicated by the color bar. The C/O ratio in a static disk (black lines) is shown for comparison. The arrows show that the C/O ratio in gas will decrease inside the H$_2$O and CO$_2$
snowlines in the viscous disk, as the relative fluxes of the desorbed icy
particles and the overall nebular gas will cause an excess of oxygen gas inside these snowlines (see Paper I for details). The presence of CO in a water ice environment rather than as pure ice moves the CO snowline significantly inward by $\sim$70\%. Taken together, disk dynamics and ice compositions move the CO snowline inward by a factor of $\sim$7.} 
\label{fig:CO_ratio}
\end{figure}

Here we use the model of Section \ref{sec:review} to estimate the movement of the CO and N$_2$ snowlines for different grain morphologies in a viscous disk. Figure \ref{fig:CO_ratio} shows the H$_2$O, CO$_2$ and CO snowline locations for particles with initial sizes $\sim0.05$ cm $\lesssim s \lesssim$ 7 m as well as estimates for the C/O ratio in gas and dust in a viscous disk, with the CO snowline calculated under different grain morphologies as noted above. We assume there is no carbon in the form of CH$_4$. The true snowline for particles that desorb outside the static snowline is the static snowline itself, hence desorbing particles with $s<0.05$ cm do not form true snowlines. If the CO binding environment is known, the CO snowline moves inward by up to $\sim$50 \% compared to a static disk for each case (pure and water dominated ices) due to disk dynamics. The full range of potential CO snowlines taking into account both ice compositions and disk dynamics span $\sim$8.7 AU to $\sim$61 AU, which is a factor of $\sim$7 difference. This implies that gas phase C/O ratios of order unity may be reached in the giant planet forming zone, and the CO snowline may be inside 10 AU for certain disk parameters.  

\begin{figure}[h!]
\centering
\includegraphics[width=0.5\textwidth]{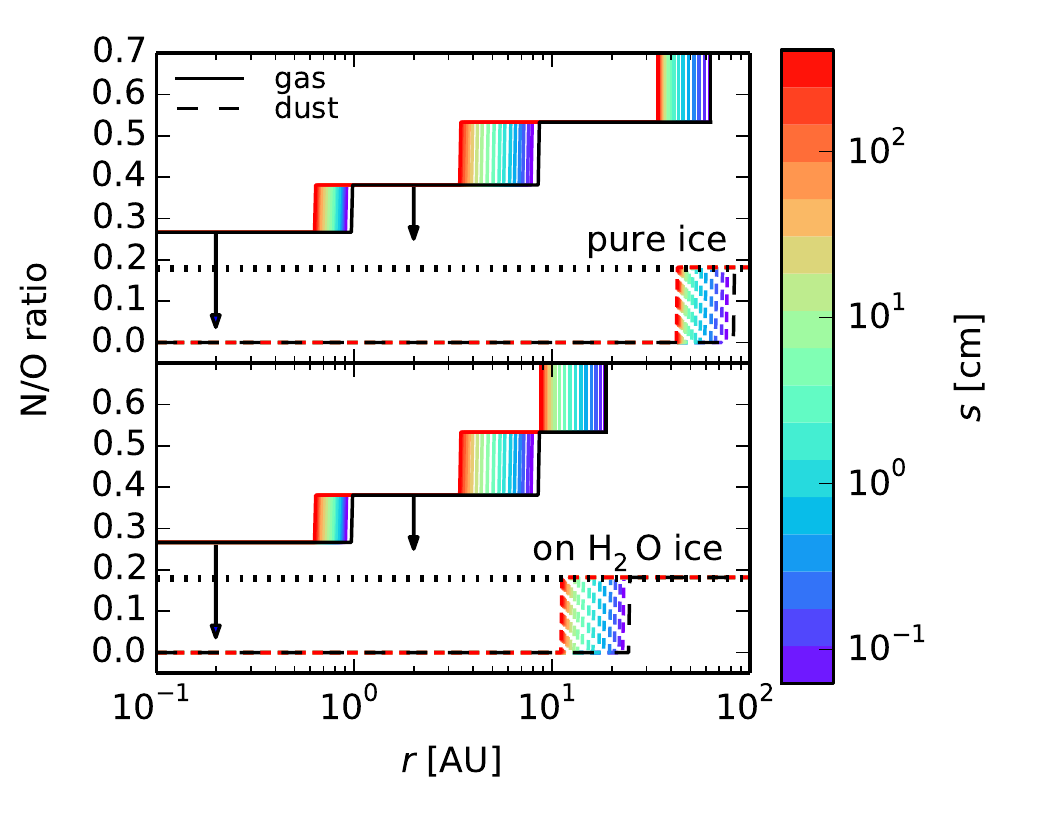}
\caption{N/O ratio estimates in gas (solid lines) and dust (dashed lines) as function of semimajor axis in a viscous disk, for CO and N$_2$ as pure ices (top panel) or as water dominated ices (bottom panel). The H$_2$O, CO$_2$, CO and N$_2$ snowlines are shown for particles with initial sizes $\sim0.05$ cm $\lesssim s \lesssim$ 7 m as indicated by the color bar. The N/O ratio in a static disk (black lines) is shown for comparison. The arrows show that the N/O ratio in gas will decrease inside the H$_2$O and CO$_2$ snowlines in the viscous disk, as the relative fluxes of the desorbed icy
particles and the overall nebular gas will cause an excess of oxygen gas inside these snowlines (see Paper I for details). Radial drift and gas accretion move the N$_2$ snowline inward by up to $\sim$50\% compared to a static disk. The presence of N$_2$ in a water ice environment rather than as pure ice moves the N$_2$ snowline significantly inward by $\sim$70\%. Taken together, disk dynamics and ice compositions move the N$_2$ snowline inward by a factor of $\sim$7. The results of an enhanced gas-phase N/O ratio between the H$_2$O and CO snowlines compared to its average value, and of highly elevated N/O ratios in gas between the CO and N$_2$ snowlines (see Figure \ref{fig:Nstatic}), are preserved.}  
\label{fig:NO_ratio}
\end{figure}

Figure \ref{fig:NO_ratio} shows the H$_2$O, CO$_2$, CO and N$_2$ snowline locations in a viscous disk for the same range of initial particle sizes as in Figure \ref{fig:CO_ratio}, and with the CO and N$_2$ snowlines calculated assuming different grain morphologies as explained above, as well as estimates for the N/O ratio throughout the disk. For simplicity, we assume that all nitrogen is the form of N$_2$. This choice is justified since the presence of some NH$_3$ only moderately changes the N/O ratio (see Figure \ref{fig:Nstatic}), and since we are primarily interested in the N$_2$ snowline locations rather than exact values for the N/O ratio. The innermost N$_2$ snowlines in the viscous disk, created by particles with $s \sim 7$ m for our fiducial model, are located at $r_{\rm N_2, pure} \approx 42$ AU for N$_2$ as pure ice and at $r_{\rm N_2, water} \approx 11$ AU for N$_2$ in water dominated ices. Thus for each case (pure versus water dominated ices), the N$_2$ snowline moves inward by up to 50\% due to disk dynamics. By taking into account both ice compositions and disk dynamics, the full range of potential N$_2$ snowlines span $\sim$11 to $\sim$79 AU, which is a factor of $\sim$7 difference. Similarly to the case for CO, the N$_2$ snowline may be close to 10 AU for certain disk models.

\section{Discussion}
\label{sec:discussion}

\begin{figure}[h!]
\centering
\includegraphics[width=0.5\textwidth]{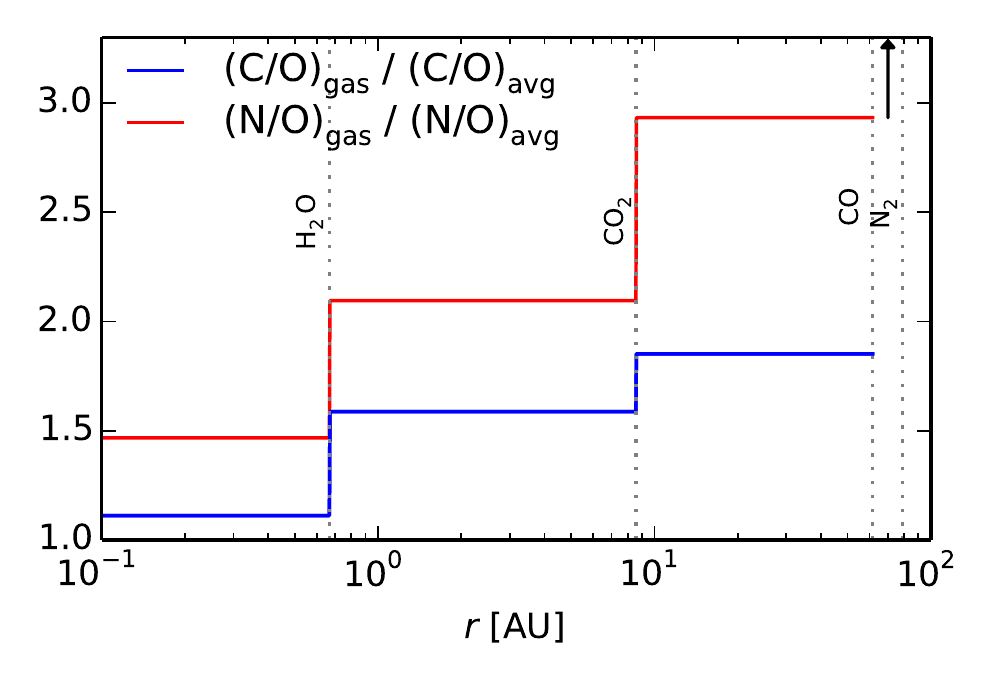}
\caption{Gas phase C/O (blue curve) and N/O (red curve) ratios divided by the average C/O and N/O ratio in a static disk, assuming CO and N$_2$ are pure ices, and there is no CH$_4$ or NH$_3$. The dashed vertical lines mark the H$_2$O, CO$_2$, CO and N$_2$ snowlines. The arrow indicates that the N/O ratio is enhanced by orders of magnitude compared to its average value between the CO and N$_2$ snowlines. The gaseous N/O ratio is enhanced throughout most of the disk, and more enhanced than the C/O ratio.}  
\label{fig:enhance}
\end{figure}

This study shows that the gas-phase N/O ratio in protoplanetary disks is considerably enhanced throughout most of the disk midplane compared to its average value. As demonstrated in Figure \ref{fig:enhance}, the gaseous N/O ratio is enhanced by a factor of two beyond the H$_2$O snowline, by more than a factor of three between the CO$_2$ and CO snowlines, and by several orders of magnitude between the CO and N$_2$ snowlines. Thus constraining the N/O ratio in a giant planet atmosphere could be used to trace its formation origins. 

Theoretical models of the magnitude and role of N/O (and N/C) ratios in exoplanet atmospheres are needed in order to use these ratios as probes for a planet's formation location. Models that explore the effect of varying the C/O ratio in exoplanet atmospheres exist in literature, and they display a large and observable effect on gas giant envelope chemistry (\citealt{lodders09}, \citealt{molliere15}). However, no similar model explorations exist for the effect of N/O and C/N/O ratios, and both are needed to exploit this potential constraint. Given the existence of such theoretical models, measurements of the N/O ratio in planetary envelopes may be possible to infer from atmospheric compositions of nitrogen versus carbon and oxygen bearing species. Nitrogen carriers have not been targeted so far due to lack of instrument sensitivity, but such observations and detections are likely in the near future with the advent of JWST (e.g., NH$_3$, \citealt{greene16}). The N/O ratio enhancement is larger than that of the gas phase C/O ratio throughout most of the disk. Thus measurements of an enhanced C/O ratio in an exoplanet atmosphere could be corroborated (disproved) by measurements of enhanced (non-enhanced) N/O ratios. Moreover, Figure \ref{fig:enhance} shows that giant planets that have formed in situ between the H$_2$O and CO snowlines are expected to present elevated both C/O and N/O ratios in their atmospheres, whereas planets between the CO and N$_2$ snowlines will have a highly enhanced N/O ratio in their atmospheres, but not C/O.  

Due to disk dynamics and ice compositions, the locations of the CO and N$_2$ snowlines, and thus the disk regions with highly elevated gas phase N/O and C/O ratios, are uncertain and may span tens of AU. Both ice morphologies discussed in this study, pure and water dominated ices, are plausible in protoplanetary disks and depend on whether H$_2$O and CO ices formed on similar timescales or successively (e.g., \citealt{garrod11}). Observations of protostellar cores show that a large fraction of CO is bound in a pure ice multilayer \citep{pontoppidan03}, but theoretical models also suggest an icy mantle structure where CO resides on a H$_2$O ice layer (e.g., \citealt{collings03}). One can also imagine a scenario where CO is in a water binding environment and N$_2$ is not. This could be attributed to the fact that H$_2$O may bind preferentially to CO than N$_2$, since both H$_2$O and CO are polar molecules while N$_2$ is not. It is also possible for N$_2$ ices to form later than CO (e.g., \citealt{pagani12}), and thus be deposited on the outer layers of the icy mantles which are typically water poor (e.g., \citealt{garrod11}). The impact of the ice environment on the snowline location is much smaller in the case of CO$_2$ and NH$_3$, as their binding energies and behavior are closer to that of H$_2$O. No detailed measurements for the CH$_4$ binding energy in a water environment exist so far, but due to its low desorption temperature a similar behavior to that of CO and N$_2$ would be expected. While the presence of some carbon in the form of CH$_4$ only modestly affects our results, CH$_4$ may become important in disks where a large fraction of the CO abundance has been converted into hydrocarbons (e.g., \citealt{du15}). 

Changes in stellar luminosity (e.g., \citealt{kennedy06}) and gas mass accretion rate (e.g., \citealt{chambers09}), as well as the evolution of icy dust particles due to grain growth and fragmentation (e.g., \citealt{birnstiel12}), may introduce additional uncertainties in the snowline locations, and thus the C/N/O ratios. Moreover, the diffusion of vapor across the snowlines following the cold finger effect (\citealt{stevenson88}, \citealt{cyr98}) will change the shape of the C/O and N/O curves and therefore the magnitude of the C/N/O ratios between different snowlines. The effect of dynamical processes on snowline locations is discussed in more detail in Paper I, Section 5.2. Given the number of uncertainties in snowline locations, detections of snowlines in a sample of disks at different evolutionary stages are needed to provide observational constraints on the relative importance of ice compositions and disk dynamics in setting snowline locations. The uncertainties in snowline locations caused by disk dynamics, ice compositions, and other effects outlined above can be resolved in extreme cases, such as a detection of a CO snowline at a temperature corresponding to pure CO ice desorption in a static disk (e.g., \citealt{qi13} at $\sim$ 17 K) or CO desorption from a water dominated ice in a dynamic disk. In intermediate cases it is more difficult to resolve the relative importance of ice compositions and disk dynamics. For example, the CO snowline in HD 163296 is at a higher temperature of $\sim$ 25 K \citep{qi15}, which could be caused either by CO being in a water dominated environment or by dynamical effects that push the CO snowline inward. Detections of multiple snowlines in the same disk could potentially break this degeneracy. 

Uncertainties in snowline locations of this magnitude also affect interpretations of Solar system observations. Recent measurements of nitrogen abundance in comet 67P/Churyumov-Gerasimenko found a N$_2$/CO ratio $\sim 10^{-3}$ \citep{rubin15}. A low N$_2$/CO ratio is consistent with comets having formed inside the N$_2$ snowline where N$_2$ is still in the gas phase. However, it is also possible that the measured N$_2$ abundance in 67P may be due to post-formation processes such as radiogenic heating \citep{rubin15}, and thus may not reflect the comet's primordial composition. Theoretical models suggest that Jupiter-family comets, such as 67P, originate from the Kuiper belt (\citealt{duncan97}; but see \citealt{rubin15} for alternative formation scenarios for 67P). It is thus possible, in principle, to use measurements of the N$_2$ abundance in Jupiter-family comets to determine where the N$_2$ snowline was located in our Solar system. However due to the uncertainty in the calculated location of the N$_2$ snowline (see Section \ref{sec:dynamic}), as well as the uncertainty of the formation zone of Jupiter-family comets (anywhere between 5 and $>$30 AU; \citealt{pontoppidan14}), more detailed modeling is needed. 


\section{Summary}
\label{sec:summary}

In this paper we explore the role of icy grain compositions and disk dynamics on the snowline locations of major volatile carrier molecules and the C/N/O ratios in protoplanetary disks. We enhance the coupled drift-desorption model developed in \citet{piso15b} by adding more carbon- and nitrogen-bearing species into our framework, and by considering different binding ice environments. Our results can be summarized as follows:

\begin{enumerate}

\item Due to the high volatility of N$_2$, the gaseous N/O ratio outside the H$_2$O snowline is enhanced by a factor of two compared to its average value, by more than a factor of three between the CO$_2$ and CO snowlines, and by many orders of magnitude between the CO and N$_2$ snowlines due to the complete depletion of oxygen gas in this region. This enhancement is more pronounced than in the case of the gas-phase C/O ratio, which is increased by at most a factor of two compared to the stellar value. 

\item The effect of CH$_4$ and NH$_3$ on the C/O and N/O ratios is small, even when we consider the maximum observed CH$_4$ and NH$_3$ abundances in protostellar cores. In this scenario, the gas phase C/O ratio increases by 8\% between the CO$_2$ and CH$_4$ snowlines, and  the gaseous N/O ratio decreases by 18\% between the NH$_3$ and CO$_2$ snowlines. In both cases, large gas phase C/O and N/O ratios in the outer disk are preserved.

\item Grain composition sensitively affects the CO and N$_2$ snowline locations. If CO and N$_2$ reside in water dominated rather than pure ices, their snowlines move inward by up to $\sim$70 \%. This effect is separate from that of radial drift and viscous gas accretion, which also cause an inward movement of the CO and N$_2$ snowlines by up to $\sim$50 \%. 

\item The locations of the CO and N$_2$ snowlines are uncertain when we consider both viscous versus static disks, and pure versus water dominated ices. The snowlines in a viscous disk with CO or N$_2$ in a water environment are by up to a factor of $\sim$7 closer to the host star that in a static disk with CO or N$_2$ as pure ices. 

\end{enumerate}

Our results have direct consequences for the composition of nascent giant planets. The considerable inward movement of the CO and N$_2$ snowlines due to the ice grains being water dominated rather than pure ices implies than giant planets with high C/O and/or N/O ratios in their atmospheres may form closer in than previously predicted by theoretical models. Moreover, our model shows that wide separation gas giants may have an excess of nitrogen in their envelopes, which may be used to trace their origins. 

\acknowledgments{We thank the anonymous referee for helpful comments and suggestions. This work is supported by a Simons Collaboration on the Origins of Life (SCOL) investigator award to KIO. JP was supported by the Banneker institute.}

\if\bibinc n
\bibliography{refs}
\fi

\if\bibinc y

\fi

\end{document}